\documentclass[%
showkeys,
showpacs,
reprint,
superscriptaddress,
nofootinbib,
nobibnotes,
 amsmath,amssymb,
aps,
prc,
]{revtex4-2}
\usepackage{amssymb}
\usepackage{graphicx}
\usepackage{dcolumn}
\usepackage{bm}
\usepackage{hyperref}
\usepackage{breakurl}

\begin{document}

\preprint{}

\title{Mathematical formulae for neutron self-shielding properties of media in an isotropic neutron field}

\author{Ateia~W.~Mahmoud}\email{atia.mahmoud@gmail.com} \email{ateia.mahmoud@eaea.org.eg}
\address{Physics Department, Faculty of Science, Ain Shams University, Cairo, Egypt.}
\address{Reactor Physics Department, Nuclear Research Center, Egyptian Atomic Energy Authority, Cairo 13759, Egypt.}

\author{Elsayed~K.~Elmaghraby}\email[Corresponding Author: ]{e.m.k.elmaghraby@gmail.com}
\email{elsayed.elmaghraby@eaea.org.eg}
\address{Experimental Nuclear Physics Department, Nuclear Research Center, Egyptian Atomic Energy Authority, Cairo 13759, Egypt.}

\author{E.~Salama}\email{e\_elsayed@sci.asu.edu.eg}
\address{Basic Science Department, Faculty of Engineering, The British University in Egypt (BUE), Cairo, Egypt.}

\author{A.~Elghazaly}\email{an\_4558@yahoo.com}
\address{Reactor Physics Department, Nuclear Research Center, Egyptian Atomic Energy Authority, Cairo 13759, Egypt.}

\author{S.~A.~El-fiki}\email{soadelfiki@sci.asu.edu.eg}
\address{Physics Department, Faculty of Science, Ain Shams University, Cairo, Egypt.}

\date{June 30, 2021}

\begin{abstract}
The complexity of the neutron transport phenomenon throws its shadows on every physical system wherever neutron is produced or used. In the current study, an \emph{ab initio} derivation of the neutron self-shielding factor to solve the problem of the decrease of the neutron flux as it penetrates into a material placed in an isotropic neutron field. We have employed the theory of steady-state neutron transport, starting from Stuart's formula. Simple formulae were derived based on the integral cross-section parameters that could be adopted by the user according to various variables, such as the neutron flux distribution and geometry of the simulation at hand. The concluded formulae of the self-shielding factors comprise an inverted sigmoid function normalized with a weight representing the ratio between the macroscopic total and scattering cross-sections of the medium. The general convex volume geometries are reduced to a set of chord lengths, while the neutron interactions probabilities within the volume are parameterized to the epithermal and thermal neutron energies. The arguments of the inverted-sigmoid function were derived from a simplified version of neutron transport formulation. Accordingly, the obtained general formulae were successful in giving the values of the experimental neutron self-shielding factor for different elements and different geometries.
\end{abstract}


\maketitle


\section{Introduction}
Over time, neutron activation analysis has been evolving into a very effective nuclear analytical technique. Such techniques are often utilized for non-destructive elemental concentration measurement in unknown materials and nuclear material interrogation \cite{TohamyElmaghrabyComsan2019162387,TohamyElmaghrabyComsan2021045304}. The constraints include neutron fluence, the fraction of fluence reaches the interior of the sample, sample mass and sample geometry \cite{TohamyElmaghrabyComsan2020109340,NAKAMURA2013119,AliandElmaghraby202063,FarinaArbocco2012,Elmaghraby2018PhysScr,Chilian2010429,Elmaghraby2019PhysScrCode,Elmaghraby2016Shape,Jacimovic2010399,Elmaghraby201742}.   Furthermore, the neutron's energy spectrum is varied, but ideally suited to research using the ideal Maxwellian distribution at room temperature, while the other distributions must be altered to match the reference nuclear reaction data \cite{Elmaghraby2018PhysScr,Elmaghraby2019NPA,TohamyElmaghrabyComsan2020109340}. Apart from that, neutrons are deeply employed in two significant geometries, including but not limited to:  (1) \textit{beam geometry}, where the neutron currents are assumed  to travel in one direction, and (2) \textit{field geometry}, where neutrons impact the sample from all directions presuming the material is isotropic. There is an important functional difference between these two geometries, i.e. the effect on the neutron flux itself. For instance, when exposing a sample to a neutron beam, the interior of the sample will be exposed to a lesser neutron fluence than the exterior part, in all circumstances regardless of the geometry of the neutron current. This phenomenally known as \emph{self-shielding}, and it is a critical element of the neutron transport phenomena. In the case of field geometry, the net neutron current essentially disappears, while the fluence (or flux) becomes the observable quantity.  There is an interplay between neutron absorption in the sample and the overall neutron flux \cite{Elmaghraby2021ICPAP2021C1}.  Predominantly, the correlation between neutrons self-shield factors and the set of parameters involved in the calculation of its value had been studied by several scientists \cite{Fleming19821263,Blaauw1995403B,Gonalves2001447,Martinho2003371,Salgado2004426,Martinho2004637,Goncalves2004186,Sudarshan2005205,Nasrabadi2007473,Mashkovich1983Book,Moll2020106990}, who gave dimensionless variables to identify and encompass the physical and geometric varieties of the samples geometries in order to attain a universal formula for self-shielding. The Mont\`{e}-Carlo approach effectively calculates self-shielding, but it takes time and an experienced user to achieve acceptable accuracy and efficiency \cite{Larsen2006513Proc}, see Appendix  \ref{App:Merit}.  Empirical expressions, such as those given by researchers in Refs. \cite{Gonalves2001447,Martinho2003371,Martinho2004637,Goncalves2004186,Salgado2004426}  based on  Ref. \cite{Fleming19821263}, had became routine in calculating self-shielding, these empirical expressions have been derived for a few specific geometries and limited number of elements.

Herein, we present a complete investigation of the neutron self-shielding phenomenon in different media. Additionally,  we had provided a full description of the physics behind the theme, taking into consideration the neutron transport inside the sample, and the absorption and scattering phenomena as a function of neutron energy. We aim to transform the problem from a spectroscopic set of parameters, usually unreachable for the common user, to an integrated set of well-known parameters and factors. The use of detailed spectroscopic parameters, such as ENDF data, cross-section, detailed dimensions and shape, the widths of neutron resonances for either scattering or absorption, etc., requires time and experienced users to make use of them with acceptable accuracy and efficiency, the cost most scientists cannot afford to just calculate a single parameter in their routine work. Our intention is focused on avoiding such cost and enhancing present existing formulae and using of an integral set of the well-known parameters, such as thermal cross-section,  resonance integral, average chord length.  All remaining factors are calculated from these three parameters. Though, existence of a mathematical formulation of self-shielding in material of different geometries and composition shall deliver additional tool to improve precision of reaction parameters and activation analysis calculations.

\section{Materials and Methods}
\label{Sec:MaterialMethods}
Experimentally measured and theoretically calculated data were collected from different sources for the self-shielding factor in In, Au, Co, Cu, and Fe samples. The geometries for these elements  were foils, wires, and infinite slabs. The Experimental data of G$_{th}$ of
Mahmoud et al. \cite{MahmoudElmaghrabySalamaElghazalyElFikiBJP2022}, Taylor \& Linacre \cite{taylor1964use}, Carre et al. \cite{carre1965etudes},
Hasnain et al. \cite{hasnain1961thermal}, Sola \cite{Sola1960},  Walker et al. \cite{walker1963thermal}, Klema \cite{klema1952thermal}, and Crane $\&$ Doerner \cite{crane1963thermal} were digitized from Martinho et al. \cite{Martinho2004637}.  The data of Eastwood \& Werner \cite{eastwood1962resonance} for Co Wire was collected from their original values.

For the epithermal neutron energy region, The experimental results from literature of Gonalves et al. \cite{Gonalves2001447}, Lopes \cite{lopes1990effect,lopes1991sensitivity}, McGarry \cite{mcgarry1964measurement}, Brose \cite{brose1964messung}, Yamamoto et al. \cite{yamamoto1965self}, Jefferies et al. \cite{jefferies1983analysis}, Eastwood $\&$ Werner \cite{eastwood1962resonance}, and Kumpf  \cite{kumpf1986self} were used.

Having two energy ranges at thermal neutron region and epithermal neutron energies, the cross-section was taken as the element-averaged thermal neutron cross-section and the resonance integrals for 1/E averaged neutron distribution as given in Table \ref{ElementCrossSections} based on previous evaluations \cite{Elmaghraby2016Shape,Elmaghraby2017Treat,Elmaghraby2019PhysScrCode}. These data shall be used in the specific calculations presented in the present work.

\begin{table*}[htb]
\caption{Element-averaged cross- section and resonance integral data. $\sigma_g$ and $\sigma_s$, are the capture and scattering cross-section at thermal energies,  $\sigma_a$,   $\sigma_t$  are the absorption and total cross-sections at thermal energies. Similarly, $I_g$, $I_s$ , $I_a$  and $I_t$  have the same respective meaning of its subscript. Data were taken from  \cite{Elmaghraby2016Shape,Elmaghraby2019PhysScrCode}. }\label{ElementCrossSections}
\centering
\begin{small}
\begin{tabular}{c|cccc|cccc}
  \hline\hline
        Isotope        & $\sigma_g$ &  $\sigma_s$ & $\sigma_a$ & $\sigma_t$ & $I_g$ & $I_s$ & $I_a$ & $I_t$ \\
              & (b)  & (b)& (b)& (b) & (b) & (b) & (b) & (b) \\
\hline
    Na    & 0.528  & 3.3929 & 0.528 & 3.9209 & 0.3021  & 130.81 & 0.3021 & 131.11 \\

    Mn    & 13.275  & 2.1163 & 13.275 & 15.391 & 13.168  & 621.33 & 13.168 & 634.5 \\
    Fe    & 2.5615  & 11.35 & 2.5615 & 13.912 & 1.2706  & 127.09 & 1.2706 & 128.36 \\

    Co    & 37.173  & 6.0319 & 37.173 & 43.204 & 74.78  & 791.53 & 74.78 & 866.31 \\

    Cu    & 3.7531  & 7.8424 & 3.7531 & 11.595 & 4.0309  & 129.89 & 4.0309 & 133.93 \\

    In    & 194.07  & 2.5686 & 194.07 & 196.64 & 3088.5  & 214.12 & 3088.5 & 3302.6 \\

    Au    & 98.672  & 7.9298 & 98.672 & 106.6 & 1567.9  & 405.52 & 1567.9 & 1973.4 \\
    \hline
    \end{tabular}
    \end{small}
\end{table*}


The uncertainty of digitized data was difficult to determine due to the use of different linear and logarithmic scales in old graphs and dependence among them. We had used the following formula $\sigma=$ $\sigma_X$+$\sigma_Y$, with $\sigma_X$ and $\sigma_Y$ are the dependent uncertainties in the digitized X and Y coordinates in the graph. The typical value of digitization uncertainty was less than 2\% which was added to the reported uncertainty, if available.

\section{Results and discussion}

According to Stuart \cite{Stuart1957617}, the  old  quantity  for an  absorbing  body  is  its  neutron blackness,
\begin{equation}\label{AEq:Blackness1}
\beta=\frac{j_{in}-j_{out}}{j_{in}},
\end{equation}
based on  the neutron current density entering the body ($j_{in}$) or going out of it ($j_{out}$).  Stuart \cite{Stuart1957617} had derived the formula for $\beta$ based on variational principle and assuming  uniform isotropic neutron field with scattering that do not change energy spectrum on the neutrons (change of energy is treated as absorption).
 Blaauw \cite{Blaauw1995403B,Blaauw1996431} began with Stuart's formula  \cite{Stuart1957617};
\begin{equation}\label{AEq:Blackness2}
  \beta=\frac{\frac{\Sigma_a}{\Sigma_t}P_0}{1-\frac{\Sigma_s}{\Sigma_t}\left(1-\frac{1}{\Sigma_t \bar{\ell}}\, P_0\right)}
\end{equation}
Here, $P_0$ is the probability of the first interaction derived from the transport kernal \cite{Blaauw1996431,Stuart1957617} in steady-state.
\begin{equation}\label{AEq:P0}
  P_0=\frac{\sigma_t}{S}\int_V \Pi_0(\vec{r}) d\vec{r},
\end{equation}
where $\Pi_0(\vec{r})$ is the unscattered flux within the material;
\begin{equation}\label{AEq:Pi0}
  \Pi_0(\vec{r})=4\int_S (\vec{n}\cdot\vec{\lambda}) \mathbb{G}(\vec{r},\vec{r}^\prime) dS,
\end{equation}
\noindent Here, $\vec{n}$ and $\vec{\lambda}$ are unit vectors in the direction of the normal to the surface and the neutron wavevector, respectively. The point symmetry neutron collision kernel has the form of Green's function \cite{Davison1957Book,Henderson1989172,Hormander2018,Moll2020106990}:
\begin{equation}\label{AEq:G}
   \mathbb{G}(\vec{r},\vec{r}^\prime)=\frac{\exp\left(-\Sigma_t |\vec{r}-\vec{r}^\prime|\right) }{4\pi |\vec{r}-\vec{r}^\prime|^2},
\end{equation}
\noindent that gives the probability a neutrons shifts between  phase-space coordinates $\vec{r}$ and $\vec{r}^\prime$ in one collision in point geometry. Time reversal applied in such cases by interchange or $\vec{r}$ and $\vec{r}^\prime$. In general, other geometries had asymptotic form as point like geometry \cite{Henderson1989172,Moll2020106990}. {The value of $\Pi_0(\vec{r})$ equals to 4 in case of nonexistence of the material in the medium because in Eq. \ref{AEq:Pi0} become unity. And that is the condition that must be satisfied by the transport kernels in all geometries.}  Multiple collision probability may be obtained through recursive relation \cite{Stuart1957617}
\begin{equation}\label{AEq:Pn}
  P_n=1-\frac{\int_V \Pi_{n-1}(\vec{r}) \Pi_n(\vec{r}) d\vec{r}}{4\int_V \Pi_{n-1}(\vec{r}) ) d\vec{r}},
\end{equation}
\begin{equation}\label{AEq:Pin}
  \Pi_n=\frac{1}{\bar{\ell}}\int_S \Pi_{n-1}(\vec{r^\prime}) \mathbb{G}(\vec{r},\vec{r}^\prime) dS,
\end{equation}
The value of $\bar{\ell}$ is the  average of Chord Length Distribution (CLD) may be weighted with the cosine of the angle between chord and the normal to the surface,
\begin{equation}\label{AEq:ellbar}
  \bar{\ell}=\frac{\int_0^{\pi} \int_0^{\pi/2} V(\underline{r},\underline{\theta},\underline{\phi}) \cos(\underline{\theta}) d\underline{\theta} d\underline{\phi}}{\int_0^{\pi} \int_0^{\pi/2} S_{\bot}(\underline{r},\underline{\theta},\underline{\phi}) \cos(\underline{\theta}) d\underline{\theta} d\underline{\phi}}.
\end{equation}
Here, $S_{\bot}$ is the surface area perpendicular to the direction of the neutron. The value of mean CLD for a convex body is related to the first Cauchy formula \cite{Bair2020arXiv200300438B} of the integration for cylindrical shape comprises 4V/S \cite{Dekruijf2003549,Trkov2009553}, where $S$ is the surface area inclosing a volume $V$ . This value may be used as the upper limit of one of the integrations. {Note that the coordinate variables were underlined in order to avoid confusion among symbols}.

There was equivalent definition of the sample blackness, the self-shielding factor denoted $G_{(\mathrm{energy~domain})}$;  which is defined as the ratio between the volume-averaged fluence rate within the material's volume that may absorb or scatter  neutrons and the fluence rate within the same volume considering absence of the interaction with neutrons.  According to  Blaauw \cite{Blaauw1995403B},
\begin{equation}\label{AEq:SeflShieldEnergyDomain0}
  G_{(\mathrm{energy~domain})}=\frac{\int_{V}\psi(\vec{r}) d\vec{r}}{V}+(\mathrm{higher~order~terms})
\end{equation}
\noindent the higher order terms were introduced by  Blaauw \cite{Blaauw1995403B} for extended neutron velocity distributions -- denoted here $\mathcal{R}$. Remembering that the $G_{(\mathrm{energy~domain})}$ is neutron energy specific parameter, any perturbation of the neutron energy distribution shall affect the experimental results as discussed in earlier work \cite{MahmoudElmaghrabySoliemanSalamaElghazalyElFikiarxiv220413246}.

\subsection{Mathematical model}
\label{ASec:Derivationformula}
In accordance to previous constraint of self-shielding formulae, we shall use  Eq. \ref{AEq:Blackness2}, include the high order in Eq. \ref{AEq:SeflShieldEnergyDomain0}, and use the relation of blackness and self-shielding \cite{Blaauw1996431} (i.e. $\Sigma_a \bar{\ell} G$=$\beta$ as first approximation). The combined formulae comprises;
  \begin{equation}\label{AEq:Blackness4}
  \begin{split}
   G_{(\mathrm{energy ~domain})} &
      =\frac{\left(\frac{\Sigma_t}{\Sigma_s}\right)}{1+ \Sigma_t\frac{\bar{\ell}}{P_0}\left(\frac{\Sigma_a}{\Sigma_s}\right) }+\mathcal{R}.
   \end{split}
\end{equation}

\noindent The value of $\mathcal{R}$ was found to has negligible contributions except when relying on the entire range of Maxwellian thermal neutron distribution, as proven in Appendix \ref{ASec:Contributionofvelocitydistribution}. However, and for the practical of constraining the thermal neutron energy range by the cadmium cutoff energy around 0.5 eV, this term can be neglected.  Hence,

  \begin{equation}\label{Eq:SelfShieldWithR}
  G_{(\mathrm{energy ~domain})}
  =\left(\frac{\Sigma_t}{\Sigma_s}\right)\frac{1}{1+ \Sigma_t\frac{\bar{\ell}}{P_0}\left(\frac{\Sigma_a}{\Sigma_s}\right) }
  \end{equation}

The value of the first term  in Eq. \ref{Eq:SelfShieldWithR} is correct when it less than 1, i.e. under the condition:
\begin{equation}\label{AEq:Conditionofexceeds1}
\left(\frac{\Sigma_t}{\Sigma_s}\right) \leq 1+ \Sigma_t\frac{\bar{\ell}}{P_0}\left(\frac{\Sigma_a}{\Sigma_s}\right)
\end{equation}
or
\begin{equation}\label{AEq:Conditionofexceeds2}
\Sigma_t \leq \Sigma_s+ \frac{\Sigma_t\bar{\ell}}{P_0}\Sigma_a
\end{equation}
i.e. the parameters in  the first term  in Eq. \ref{Eq:SelfShieldWithR} must satisfy the condition.
\begin{equation}\label{AEq:Conditionofexceeds3}
\frac{\Sigma_t\bar{\ell}}{P_0} \geq 1
\end{equation}
Precise choice of the value of $P_0$ is given in Section \ref{sec:P0}.

Eq. \ref{Eq:SelfShieldWithR} can be rewritten as follows:
\begin{equation}\label{Eq:Selfshielding}
  G_{(\mathrm{energy ~domain})} =\left(\frac{\Sigma_t}{\Sigma_s}\right)\times\frac{1}{1+\mathcal{Z}}
\end{equation}
\noindent where
\begin{equation}\label{Eq:dimlesspar}
\mathcal{Z}=\underbrace{\Omega(\bar{\ell},\Sigma_a,\Sigma_s,\Sigma_t)}_\mathrm{Geometry}\underbrace{\chi(\Sigma_t)}_\mathrm{Composition}\underbrace{\eta(\Sigma_a,\Sigma_s)}_\mathrm{Probability}.
\end{equation}
\noindent  The factor $\left(\frac{\Sigma_t}{\Sigma_s}\right)$  represents the weight of the total interaction cross-section versus the scattering contribution. The dimensionless parameter, $\mathcal{Z}$,
is expressed as product of three functions, geometry function ($\Omega=\cfrac{\bar{\ell}}{P_0}$) as a function of the dimensions of the sample in the unit of [cm], macroscopic cross-section function ($\chi=\Sigma_t$) in the units of [cm$^{-1}$] which depends on the isotopic content of the sample, and a dimensionless neutron energy correcting factor ($\eta=\cfrac{\Sigma_a}{\Sigma_s}$) which is a function of the neutron absorption  and the scattering cross-sections.
The neutron-chord length is not the only parameter in the transport equation that depends on geometry; the first and higher order interaction probabilities, i.e. $P_0$ , are also dependent on both geometry and medium contents. These are the fundamental morphological descriptors of the media that describes the mean intercept length and relative to the mean free-paths of neutron within the medium. Here, $\Omega$  contains the factor of a shift in the Euclidean distance value and reflecting the total distance of interest within the medium, while $\chi$ and $\eta$ determine the slope of the steeping part of the curve.

Macroscopic cross-section function is expressed as
\begin{equation}\label{Eq:MacroscopicXSelemental}
 \chi(E_n) =\Sigma(E_1,E_2) = \frac{\rho N_A \theta_i}{M}  \tilde{\sigma} ,
\end{equation}
\begin{equation}\label{Eq:sigmatild}
\tilde{\sigma}=\frac{\int_{E_1}^{E_2}\sigma_{c,\,i}(E_n)\varphi(E_n)dE_n}{\int_{E_1}^{E_2}\varphi(E_n)dE_n}
\end{equation}
\noindent where $N_A$ is the Avogadro's number [mol$^{-1}$], $\rho$ is the density of the material [g cm$^{-3}$], and $M$ is its atomic mass [g mol$^{-1}$]. $\theta_i$ is the isotopic abundance of the absorbing isotope in which it should be multiplied  by the fraction of the element in the material if a compound material is used.  Here, $\tilde{\sigma}$ is the integral cross-section in the \emph{energy domain} between $E_1$ and $E_2$. For practical purposes, the thermal neutron energy range is bounded by the cadmium cutoff energy around 0.5 eV while the epithermal range extends from 0.5 to few MeVs  \cite{TohamyElmaghrabyComsan2020109340,Elmaghraby2016Shape,Elmaghraby2018PhysScr}. Here, $\sigma_{c_i}(E_n)$ is the $i^{th}$-isotope's cross-section for the reaction channel $c$ at the neutron energy $E_n$ in this energy domain [cm$^2$]. In case of compounds, this formula becomes a summation over $i^{th}$-isotope.

 The $\eta$ term is the scattering to absorption ratio;
\begin{equation}\label{Eq:ScatteringTerm}
  \eta(\Sigma_s,\Sigma_a)=\frac{\Sigma_a}{\Sigma_s}
\end{equation}
{In general, values of thermal cross-section and resonance integral are well known from tables \cite{Elmaghraby2016Shape,Sukhoruchkin1998book,Sukhoruchkin2009NeutResPara} or integration of spectroscopic cross-section \cite{Elmaghraby2019PhysScrCode}}.

\subsection{The geometry factor}
\label{Sec:GeometryFactor}
{The geometry factor depends on the neutron-chord length and the probability of interaction as;
\begin{equation}\label{Eq:GeometryFactor}
  \Omega(\text{shape parameters})=\frac{\bar{\ell}}{P_0}.
\end{equation}
There were great efforts to parameterize this factor through years. Recent efforts had been made by  Trkov et al. \cite{Trkov2009553} especially in the extended range of the neutron resonances}.

In integral geometry, obtaining the orientation-dependent chord lengths of a convex body, in general, is a complicated mathematical argument; most researchers treat the problem with the body that has the minimal volume in a class of convex bodies having the same dimensions \cite{Horvath20201}. As long as we want to escape the rigorous derivations of the  average neutron-chord length (cf. Refs. \cite{Mazzolo20031391,Roberts19994953,Dekruijf2003549,Zoia201920006,Khaldi201713,Zhang1999985} for details), we shall use simple formulation of the average neutron-chord length based on the fact that the trajectories of the incident isotropic neutrons traverse different lengths within the body due to scattering. The derivation proceeded under this assumption in which the convex bodies are in an isotropic neutron field are the geometry which the most materials have.

\begin{figure}[htb]
  \centering
  \includegraphics[width=\linewidth]{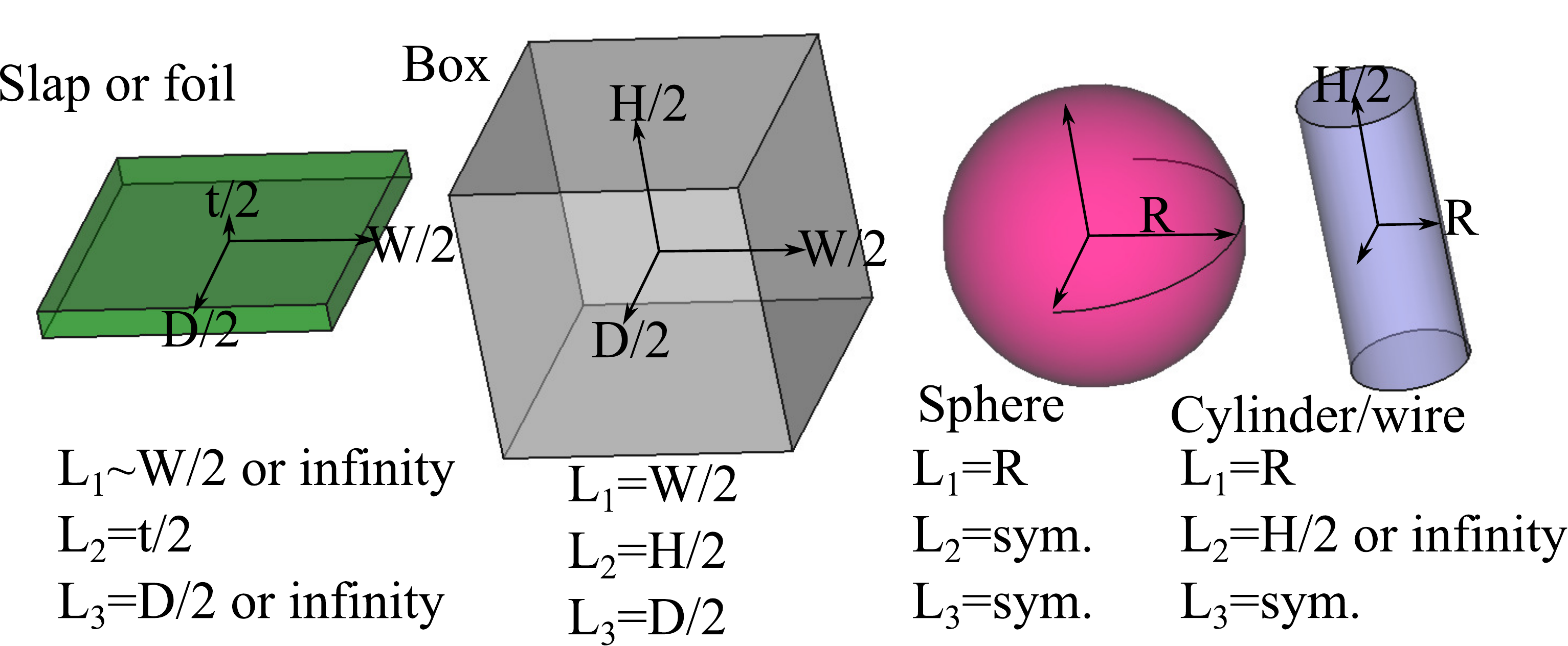}
  \caption{(Color on-line) Most known shapes of neutron activated materials. $L_1$, $L_2$, and $L_3$ are the measurable lengths, see text.}\label{Fig:Shaps}
\end{figure}

Considering an Euclidean space ($\mathbb{E}^3$) where the irradiated body is located with a center-of-mass  at the origin of the coordinate system. The three coordinate vectors are in orthogonal directions -- denoted $\hat{1}$, $\hat{2}$, and $\hat{3}$. The orientations of these coordinates were chosen as follows:  At least one of these vectors ($\hat{1}$) shall intersects the body surface in direction of the shortest length between the center-of-mass (at the $\mathbb{E}^3$ origin) and a point on the surface of the body, see Fig. \ref{Fig:Shaps}.
The next coordinate vector ($\hat{2}$) shall lay in a plan perpendicular to the first one to the shortest point on the body surface. The third coordinate vector ($\hat{3}$) shall be perpendicular to the plan  containing the first and second vectors and had length equals to the distance to the surface. The lengths of the distances between the center-of-mass and the actual surfaces along the coordinate vectors {are denoted $L_1$, $L_2$, and $L_3$, in their respective order}. Due to scattering, the coordinates are transformed to another \textit{virtual} coordinate system that is suitable to the situation and the shape on hand. The distances travelled by the neutron along the new virtual coordinate system in the body are the neutron-chord lengths, denoted $\ell_1$, $\ell_2$, and  $\ell_3$, which need to be determined using transport equations. The average neutron-chord length is taken, in the present work, as the harmonic mean of these three distances; i.e.
\begin{equation}\label{Eq:Averageell}
  \frac{1}{\bar{\ell}}=\frac{1}{3}\left(\frac{1}{\ell_1}+\frac{1}{\ell_2}+\frac{1}{\ell_3}\right).
\end{equation}
\noindent Here, $\bar{\ell}$ is the average neutron-chord length.

Due to symmetry operations in the diffusion equation, we shall take the condition that if there were a symmetry making one or more of these virtual lengths undeterminable, it should take an infinity value.
The behavior of the absorption and scattering transport are strongly dependent on each other and depend on the neutron energy \cite{MahmoudElmaghrabySoliemanSalamaElghazalyElFikiarxiv220413246}. However, a single group that cover the entire energy range of neutron (both thermal and epi-thermal) can be used to obtain meaningful value of the average neutron-chord length. The time dependent diffusion equation comprises;
\begin{equation}\label{Eq:TimeDepDiffusion}
  \frac{1}{{\mathrm{v}_{av}}}\frac{\partial \varphi (\vec{r},t)}{\partial t}=\nabla \cdot J(\vec{r},t)-{{\Sigma }_{a}}\varphi (\vec{r},t)+Q(\vec{r},t),
\end{equation}
\noindent where {$\varphi(\vec{r},t)$ is the flux}{,}  $J(\vec{r},t)=$ $D(\vec{r},t)$$\nabla \varphi (\vec{r},t)$  is the neutron current,  $Q(\vec{r},t)$ is the neutron production rate within the medium in units of [n~cm$^{-3}$s$^{-1}$]. {The sample, however, is embedded, presumably, within a uniform neutron field in which the flux outside it, denoted $\varphi_o$,  is isotropic, uniform and does not depend on the diffusion within the sample.  Considering our situation of sample absorbing neutrons, the solution of the problem comes as difference-problem in the steady-state where  $Q(\vec{r},t)$ equated to $\varphi_o$,  and considering only the difference within the sample;
\begin{equation}\label{Eq:varphitophi}
\varphi(\vec{r},t)=\varphi_o-\phi(\vec{r},t),
\end{equation}
\noindent Under the condition of steady-state, the time derivative vanishes; while Eq. \ref{Eq:TimeDepDiffusion} is reduced to:
\begin{eqnarray}
 \label{Eq:Diffusion} \nabla\cdot D(\vec{r})\nabla \phi(\vec{r})\text{-}{{\Sigma }_{a}}\phi  (\vec{r},t)&=0 &  \text{ inside the sample,}\\
\label{Eq:Diffusion1} \phi  (\vec{r},t)& =0 &  \text{ outside the sample.}
\end{eqnarray}
In the homogenous isotropic medium, $D(\vec{r})$ and $\Sigma_a(\vec{r})$ are constants, so Eq. \ref{Eq:Diffusion} can be rewritten as:}.
\begin{equation}\label{Eq:leakagerate1}
    \nabla^2\phi(\vec{r})-B^2\phi(\vec{r})=0,
\end{equation}
The $B^2$ factor depends on body materials usually called the \emph{geometric buckling factor} {in units of [cm$^{-2}$s$^{-1}$], see Appendix \ref{AApp:Derivation}.  In steady-state, and in the present work, we rewrite the factor $B^2$  as {reciprocal squared dimensions multiplied by (1 s$^{-1}$)}{};
\begin{equation}\label{Eq:B2}
  B^2=\sum_{i=1}^{3}B_i=\sum_{i=1}^{3} \frac{{\pi}^2}{\ell_i^2} (1\, s^{-1}),
\end{equation}
\noindent where the values of $\ell_i$ are the virtual dimensions of the sample in the transformed coordinates.

The diffusion length is the mean square distance that a neutron travels in the one direction from the source to its absorption point. The steady-state condition requires that the neutron current through the body's surfaces at its measurable dimensions, denoted $L_1$, $L_2$ up to $L_3$, be constant, {i.e. the change in the  leakage of current ($\left.\nabla \cdot J(\vec{r})\right|_{\vec{r}_B}$)  at these boundaries, $\vec{r}_B$, be zero}. For convex surface, a neutron leaving the region through the surface cannot intersect the surface again. Consequently,  general solution {at the surface} {shall} satisfy the condition{s}{;}
\begin{eqnarray}
\nonumber \label{Eq:CondittionForell1} \left.\nabla \cdot D(\vec{r})\nabla \phi(\vec{r})\right|_{\vec{r}_B}&\equiv&\left.\nabla \cdot J(\vec{r})\right|_{\vec{r}_B}=0,\\
   \label{Eq:CondittionForell2} J(\vec{r}>\vec{r}_B)&=&J(\vec{r}\leq\vec{r}_B).
\end{eqnarray}
{Which refer to the continuity of flux at the surface and constancy of current, respectively. Under this condition, the boundary in the difference-problem of Eqs. \ref{Eq:Diffusion} and \ref{Eq:Diffusion1} is considered a \emph{vacuum boundary} \cite{AlmenasAndLee1992book}. At the specific boundary vector $\vec{r}_B$, the flux become $\varphi(\vec{r}_B)=\varphi_o$, or, according to Eq. \ref{Eq:varphitophi},
\begin{equation}\label{Eq:CondittionForell}
\phi(\vec{r}_B)=0
\end{equation}
}{}
The solution of Eq. \ref{Eq:leakagerate1}, as derived in Appendix \ref{AApp:DerivationCartCoor}  for cartesian coordinates, resembles;
\begin{equation}\label{Eq:Leakagerate2}
   \phi(\underline{x},\underline{y},\underline{z})=\mathrm{Const.} e^{-iB_1\,\underline{x}-iB_2\,\underline{y}-iB_3\,\underline{z}}
\end{equation}
{Note that the coordinate variables are underlined from now one in order to avoid confusion among symbols}. In order to satisfy the condition Eq. \ref{Eq:CondittionForell}, the values of $\ell_i$ should be related to  the measurable coordinate dimensions as follows: $\ell_1 =2 L_1$, $\ell_2 =2 L_2$ and , $\ell_3 =2 L_3$. As presented in Fig. \ref{Fig:Shaps}, $L_1=W/2$, similarly all other dimensions.  i.e.
\begin{equation}\label{Eq:ellBOx}
  \bar{\ell}=3\left(\frac{1}{W}+\frac{1}{D}+\frac{1}{H}\right)^{-1}
\end{equation}
Note that for infinite foils and sheets, there is only one measurable length exits, the thickness with $L_1$$=t/2$, $L_2=\infty$, and $L_3=\infty$ making $\ell_1$$=t$, $\ell_2=\infty$, $\ell_3=\infty$, and
\begin{equation}\label{Eq:ellinfinitesheet}
  \bar{\ell}=3t
\end{equation}
For cylindrical geometries the solution of  Eq. \ref{Eq:leakagerate1}, see Appendix \ref{AApp:DerivationCylinCoor}, becomes;
\noindent \begin{eqnarray}\label{Eq:leakagerate11}
\nonumber \phi(\underline{\rho},\underline{\phi},\underline{z}) = & \sum_{m=0}^{\infty}\sum_{n=0}^{\infty}C_{mn}
 J_m (\sqrt{n^2+B_1^2}\underline{\rho})  \times & \\
 \nonumber & \cos{m\underline{\phi}} e^{-\sqrt{n^2-B_2^2}\underline{z}}  \text{,  if } \, n^2>B_2^2, \\
\nonumber    =&\sum_{m=0}^{\infty}\sum_{n=0}^{\infty}C_{mn} J_m (\sqrt{n^2+B_1^2}\underline{\rho})  \times & \\ & \cos{m\underline{\phi}} \cos{\sqrt{B_2^2-n^2}\underline{z}}  \text{,  if } \,  n^2\leq B_2^2
\end{eqnarray}

\noindent where $J_m$ is Bessel functions of the first kind, $C_{mn}$ is a constant and m and n are integers. Generally, m=0 and n=0 have the largest contribution. Hence
\begin{equation}\label{Eq:leakagerate12}
  \phi(\underline{\rho},\underline{\phi},\underline{z})\simeq C_{00} J_0 (B_1\,\underline{\rho})\cos{B_2\,\underline{\phi}}
\end{equation}
\noindent  The first root of  $J_0 $ in Eq. \ref{Eq:leakagerate12} is when the argument  $B_1\underline{\rho}=$2.4048 while the root of  the cosine function is when its argument  $B_2\,\underline{\phi}$=$\frac{\pi}{2}$. In order to satisfy the condition Eq. \ref{Eq:CondittionForell};

  \begin{eqnarray}\label{Eq:L_xyzCylind}
  \ell_1 & =\frac{\pi}{2.4048} L_1 &= \frac{\pi}{2.4048}R \\
    \ell_2 & =2 L_2 &=H,
  \end{eqnarray}
  \noindent  and none-existence of  $\underline{z}$ dependance Eq. \ref{Eq:leakagerate12} gives $\ell_3=\infty$. i.e.
\begin{equation}\label{Eq:ellfinitcylinder-reduced}
  \bar{\ell}=\frac{3\pi R H}{2.4048 H +\pi R}
\end{equation}
Note that for finite disk of the radius $R$ and the thickness $t$.
\begin{equation}\label{Eq:ellinfinitedisk}
  \bar{\ell}=\frac{3\pi R t}{2.4048 t +\pi R}
\end{equation}
For  infinite wire and cylinders, there is only one measurable coordinate length, the radius $R$
\begin{equation}\label{Eq:ellinfinitewire}
  \bar{\ell}=\frac{3\pi R}{2.4048}=3.9191R
\end{equation}

The solution of Eq. \ref{Eq:leakagerate1}, as derived in the Appendix \ref{AApp:DerivationSpherCoor}, for spherical shape resembles;
\begin{equation}\label{Eq:leakagerate19}
\phi(\underline{r},\underline{\theta},\underline{\phi})= \sum_{l=0}^{\infty}\sum_{m=0}^{\infty}C_{mn} j_l(B_1\underline{r})  P_l^m(\cos\underline{\theta}) \cos{m\underline{\phi}},
\end{equation}
\noindent  where, $P_l^m(\cos\underline{\theta})$ is the associated Legendre polynomial,  $j_l(B \,\underline{r})$ is the spherical Bessel function, and  $C_{mn}$ is the integration constant. Again $C_{00}$ has largest contribution. Hence;
\begin{equation}\label{Eq:leakagerate20}
  \phi(\underline{r},\underline{\theta},\underline{\phi})\simeq C_{00} \;j_0(B_1 \,\underline{r}) \; P_0^0(\cos\underline{\theta})
\end{equation}
The condition in Eq. \ref{Eq:CondittionForell} is satisfied if the argument of the spherical Bessel function ($B_1 \,\underline{r}$) equal the root of the spherical Bessel function at $\pi$; i.e.
the condition is satisfied if $\ell_1= L_1= R$.
\noindent {Here, $P_0^0(\cos\underline{\theta}) $=1 and due to} {symmetry} of the body, there exists   none-$\underline{\theta}$  dependance which reveal $\ell_2=\infty$, while none-$\underline{\phi}$ dependance requires $\ell_3=\infty$; i.e.
\begin{equation}\label{Eq:ellsphere}
  \bar{\ell}=3 R
\end{equation}

For any other irregular shape the average neutron-chord length can be calculated in the same manner using box geometry as approximation. Table \ref{Tab:Comparisonofl} showed a comparison between our simple procedure and others.

\begin{table}[htb]
  \caption{The geometry factor $\Omega(\bar{\ell},\Sigma_t)$ deduced from simple convex geometries presented in Fig. \ref{Fig:Shaps} in comparison to reported factors from different researches. Reported factors are from Refs. Ref. \cite{Gonalves2001447,Martinho2003371,Martinho2004637,Goncalves2004186}. }\label{Tab:Comparisonofl}
  \centering
  \begin{small}
  \begin{tabular}{c | c | c | c  }
     \hline      \hline
 \multicolumn{2}{c|}{Body} &\multicolumn{2}{c}{The geometry factor $\Omega(\bar{\ell},\Sigma_t)$}\\
 \hline
  Shape & Dimensions & \multicolumn{1}{c|}{Present work} & Reported\\
  \hline
     Sheet-infinite & $t$ &${3t}/{P_0}$ &$1.5 t$ \\
     Disc & $t$,$R$ &  $\frac{3\pi tR}{(P_0(2.4048t+\pi R))}$ & --  \\
     Box/Slap & $H$, $W$, $D$ & $\frac{3HDW}{(P_0(HW+HD+DW))}$ & --\\
     Sphere & $R$& ${3R}/{P_0}$  &  $R$\\
     Cylinder & $H$, $R$& $\frac{3\pi R H}{(P_0(2.4048 H +\pi R))}$ & $1.65\frac{HR}{H+R}$\\
    Cylinder-infinite & $R$ &   ${3.9191R}/{P_0}$ &--\\
     General shape & $L_1$, $L_2$, $L_3$ &  $\frac{3}{P_0}\left(\frac{1}{\ell_1}+\frac{1}{\ell_2}+\frac{1}{\ell_3}\right)^{-1}$ &--\\
     \hline
   \end{tabular}
  \end{small}
  \end{table}

The average neutron-chord length is larger than the dimensions of the body due to the irregular path of neutrons in the body's material.  For a sphere with radius R is $\bar{\ell}=$3R, not the value of 2R, while for infinite foil it is also three times its thickness, not the value of 1.5t, due to the average of the cosine in the neutron scattering path length inside the volume. However, when  Sjostrand et al. \cite{Sjostrand20021607} calculated the average neutron-chord length for a sphere assuming an isotropic flux distribution, the result was equal to the radius R due to use of different weighting factors.

\subsection{Probability of the neutron interaction}
\label{sec:P0}

The next step is to obtain a mathematical formula for the probability of single interaction within the volume ($P_0$), i.e. the probability that a neutron will suffer at least one more interaction. In the case of thermal energies, the domain of Maxwellian distribution below cadmium cut-off energy may be considered as that of the averaged energy at 0.025 eV for which scattering and multiple scattering shall not disturbs the overall neutron energy distribution \cite{beckurtsWirtz1964}. The neutron absorptions are the result of the various neutron resonances which are predominant in the epi-thermal region including those of capture and possible fission components while scattering components cause escape of neutrons from this region.

The neutron-escape probability, the factor $p$ in reactor physics, measures the fraction of neutrons that have escaped absorption and still exist after having been slowed-down from their epi-thermal energies to -- say thermal energies -- due to these ``resonance traps'' and reduces the absorption losses \cite{Liverhant1960book,Rothenstein1960162}. Several authors had tried calculate the resonance escape probabilities from first principles \cite{Leoncini201822,Leslie196578,Christy1992BookChapter,Rothenstein1960162} while other calculate directly from thermal utilization factor of reactors (\textit{cf.} Ref. \cite{Laramore2018255}).

In the present work  and under the condition in  Eq. \ref{AEq:Conditionofexceeds3}, the probability of interaction is obtained from the  attenuation relation ($\psi=$ $\exp$ $(- \Sigma_t$ $\times$ $\mathrm{mean~distance})$). The scattered neutron continues to exist within the body. The mean travelled distance is the averaged neutron-chord length which allows us to write directly and according to Rothenstein \cite{Rothenstein1960162}, and for approximation, the following
\begin{equation}\label{Eq:ApproximateP0}
  P_0\sim(1-\exp(-\Sigma_t \bar{\ell})),
\end{equation}
which satisfies the condition in Eq. \ref{AEq:Conditionofexceeds3}.
In the extended range of epi-thermal neutrons, the flux varies as 1/$E_n$ from cadmium cut-off energy at 0.5 eV to the end of the neutron spectrum -- say 1 MeV. Multiple scattering disturbs the energy distribution by reducing the number of neutrons in the epi-thermal region. In reactor physics, this phenomenon is described by resonance escape probability, which is the probability that a neutron will slow down from fission energy to thermal energies without being captured by a nuclear resonance. This phenomenon depends on the diffusion properties of the medium. In the comparison given in Fig. \ref{Fig.:Epithermal}, the values of $G_{epi}$ vary as in Eq. \ref{Eq:ApproximateP0} but with $\bar{\ell}^{2/3}$ replaces $\bar{\ell}$. {There was no clear reason about this dependence. However, there is a simple experimental remark  in neutron physics: whenever the neutron energy distribution repudiate the proper thermalization distribution (Maxwellian+1/E dependance in epi-thermal region) by any mean such as absorption, the neutrons rapidly redistribute its velocity population within the diffusion distance to follow the proper distribution -- up to  thermalization. \cite{Todorov2017194202,Elmaghraby2019NIMASamar1,TohamyElmaghrabyComsan2021045304}}. {To compensate such dependence,} in the present work we had introduced a parameterized factor to enhance formula in Eq. \ref{Eq:ApproximateP0} in the epi-thermal range as follows;
\begin{equation}\label{Eq:ApproximateP0epi-thermal}
  P_0\sim \frac{1}{p_{\mathrm{escape}}}\left(1-\exp\left(-\Sigma_t \bar{\ell}{p_{\mathrm{escape}}}\right)\right),
\end{equation}
\noindent where
\begin{equation}\label{Eq:ApproximateP0epi-thermalfactor}
  {p_{\mathrm{escape}}}= 2\frac{\sqrt[2]{|\Sigma_a-\Sigma_s|}}{\sqrt[2]{\Sigma_a}\sqrt[3]{\Sigma_s\bar{\ell}}}.
\end{equation}
which, also, satisfies the condition in Eq. \ref{AEq:Conditionofexceeds3}.

The subscript $(\mathrm{energy~domain})$ is to replaced by ``$th$'' in case of thermal neutron energies below cadmium cutoff energy ($\sim$0.5 eV) or by ``$epi$'' for epi-thermal neutrons having energy domain greater than the cadmium cutoff energy.
Here, we have used two notions of flux $\varphi_o$ and $\varphi$ which stand for unperturbed neutron flux for which the material is diluted or absent \cite{Elmaghraby2021ICPAP2021C1}  and the measured self-shielded neutron flux in the vicinity of the material, respectively.

\subsection{Verification with experiment}

The obtained mathematical values with present \textit{ab initio} model were compared with experimental values in Fig. \ref{Fig:ComparisonWithExperiments}. The exact parameters of the experimental data such as foil thickness and wire radius, cylinder hight were obtained from the original sources (whether these were literature or our previous experiments). In the thermal energy range,  the  derived formula in Eqs. \ref{Eq:Selfshielding} and \ref{Eq:ApproximateP0} gave a good representation of the experimental data for In, Au and Co within the experimental uncertainty, whether it was wires or foils .
\begin{figure}[htb]
  \includegraphics[width=\linewidth]{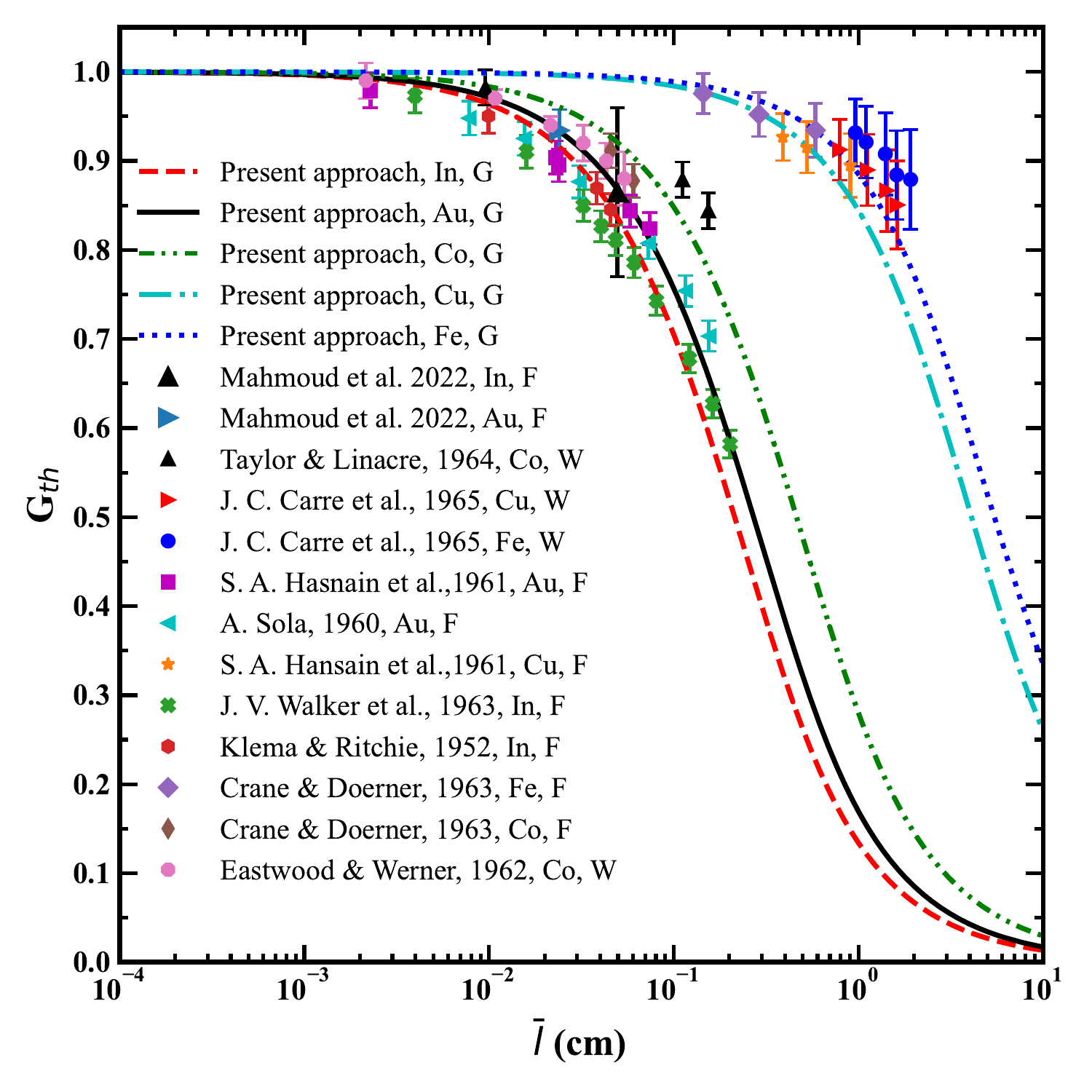}
  \caption{(Color on-line) Comparison of G$_{th}$ of our approach with Experimental values taken from the literature.  Experimental data of were those of Mahmoud et al. \cite{MahmoudElmaghrabySalamaElghazalyElFikiBJP2022}, Taylor $\&$ Linacre \cite{taylor1964use}, Carre et al. \cite{carre1965etudes}, Hasnain et al. \cite{hasnain1961thermal}, Sola \cite{Sola1960},  Walker et al. \cite{walker1963thermal}, Klema \cite{klema1952thermal}, and Crane $\&$ Doerner \cite{crane1963thermal} as adopted from Martinho et al. \cite{Martinho2004637}. The uncertainty of G$_{th}$  was added as 10\% for all experimental values of G$_{th}$ due to digitization uncertainty. (F: Foil, C: Cylinder, W: Wire, and \textbf{G}: General Convex body, our approach). Error bars are either the digitization errors or a given uncertainty, see Section \ref{Sec:MaterialMethods}.} \label{Fig:ComparisonWithExperiments}
\end{figure}

\begin{figure}[htb]
	\includegraphics[width=\linewidth]{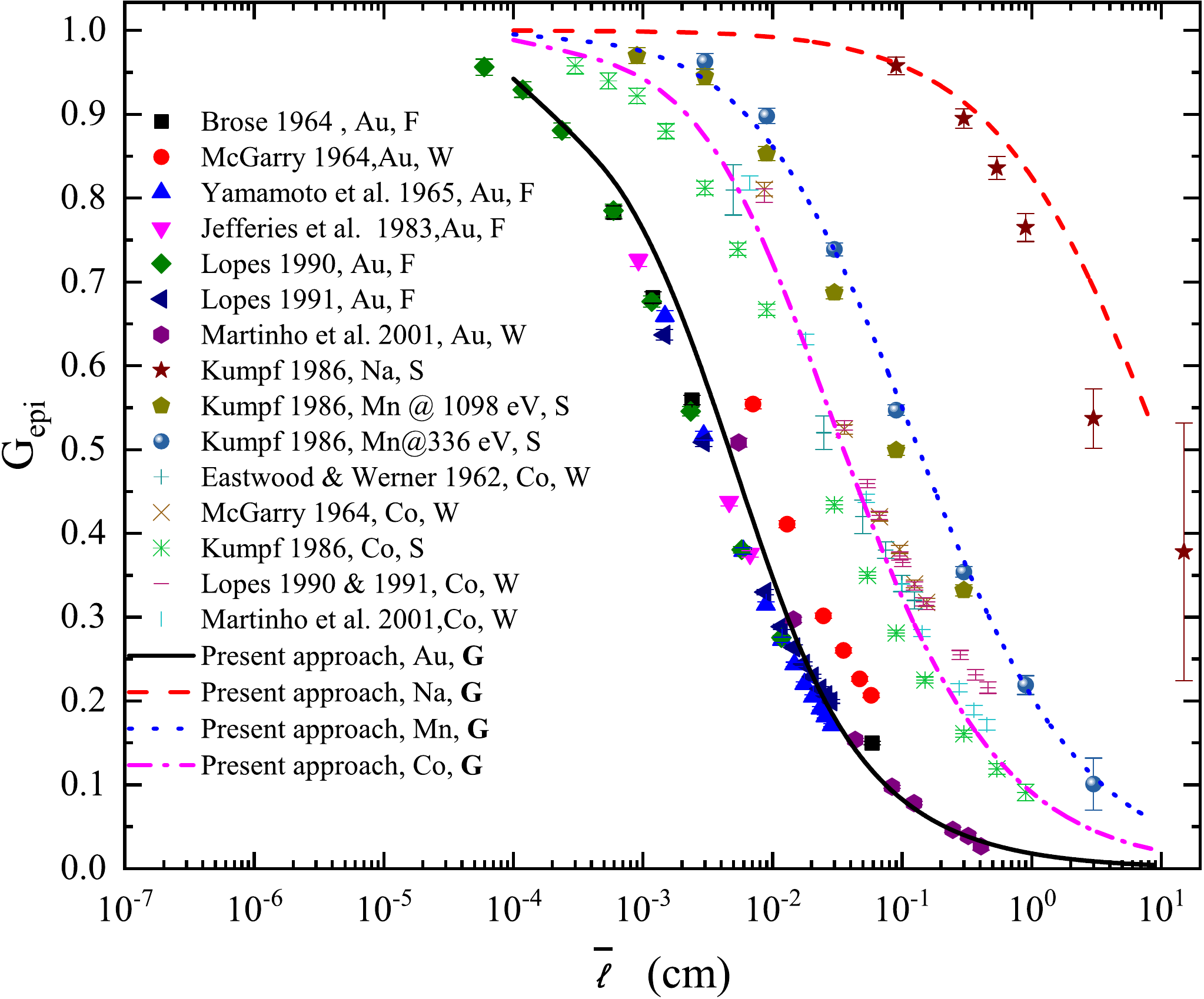}
	\caption{(Color on-line) Comparison of G$_{epi}$ of our derived formula, as in Eq. \ref{Eq:Selfshielding} using the interaction probability of Eq.  \ref{Eq:ApproximateP0epi-thermal}, with experimental values taken from  Gonalves et al. \cite{Gonalves2001447}, Lopes \cite{lopes1990effect,lopes1991sensitivity}, McGarry \cite{mcgarry1964measurement}, Brose \cite{brose1964messung}, Yamamoto et al. \cite{yamamoto1965self}, Jefferies et al. \cite{jefferies1983analysis}, Eastwood $\&$ Werner \cite{eastwood1962resonance}, and Kumpf  \cite{kumpf1986self}. (F:Foil, W:Wire, S: Infinite Slab, and \textbf{G}: General Convex body, our approach). Error bars are either given or due to digitization, see Section \ref{Sec:MaterialMethods}.}\label{Fig.:Epithermal}
\end{figure}

In the epi-thermal region, the integral cross-section of Eq. \ref{Eq:sigmatild} is replaced by the resonance integral. Table \ref{ElementCrossSections} contains the element-averaged  resonance integrals for 1/E averaged neutron distribution together with the thermal  cross- section data based on Maxwellian distribution of neutron energies for capture reactions \cite{Elmaghraby2016Shape} and scattering reactions  \cite{Elmaghraby2019PhysScrCode}.  These integral data were used to calculate epi-thermal self-shielding  factor, G$_{epi}$, and represented by lines in Fig. \ref{Fig.:Epithermal}. Experimental results from literature of Gonalves et al. \cite{Gonalves2001447}, Lopes \cite{lopes1990effect,lopes1991sensitivity}, McGarry \cite{mcgarry1964measurement}, Brose \cite{brose1964messung}, Yamamoto et al. \cite{yamamoto1965self}, Jefferies et al. \cite{jefferies1983analysis}, Eastwood $\&$ Werner \cite{eastwood1962resonance}, and Kumpf  \cite{kumpf1986self} were used for comparison.  Elements were Au, Co, Mn and Na in a form of wire, foils or infinite slabs. Note that our model calculations were based on the  derived formula in Eq. \ref{Eq:Selfshielding} and the interaction probability of Eq.  \ref{Eq:ApproximateP0epi-thermal}. Parameters of such as foil thickness and wire radius, cylinder hight were obtained from the original sources of the experimental data.  With the adaptation in the Eq. \ref{Eq:ApproximateP0epi-thermalfactor}, our model gave a good representation of the experimental data for Au, Co, Mn and Na within the experimental uncertainty. Otherwise, the model curve shall be more steeper and messes the experimental data.  As shown in Fig. \ref{Fig.:Epithermal}, there is a slightly different between experimental, calculated values and our model in gold and Indium wires and foils.

It is clear that the neutrons self-shielding  factor depends not only on the properties and geometry of the material but also on the neutron energy range as shown in Fig. \ref{Fig:ComparisonWithExperiments} and \ref{Fig.:Epithermal}. A comparison between the present approach of \emph{ab initio} calculations, considering the extreme cases of infinite wire and infinite foil, and the empirical equation given in Refs. \cite{Gonalves2001447,Martinho2003371,Martinho2004637,Goncalves2004186,Salgado2004426} are given in Appendix \ref{Asec:Comparison} of the present work.

\section{Conclusion}
{In the vicinity of neutron-absorbing elements within the sample, neutron flux shall be modified continuously with the depth. The activation formulae that take the flux as a constant value shall be corrected by a self-shielding factor. The self-shielding corrected neutron flux factor are often obtained from numerous approaches,  both empirical based on fitting or analytic analysis as presented within the present work. Understanding the physics behind self-shielding enabled the extension of a simple thermal neutron picture into the epithermal energies with the possibility for application to high-energy neutrons. Equation \ref{Eq:Selfshielding} together with its descriptive parameters in Eqs. \ref{Eq:MacroscopicXSelemental}, \ref{Eq:ScatteringTerm}, \ref{Eq:GeometryFactor}, \ref{Eq:ApproximateP0}, and \ref{Eq:ApproximateP0epi-thermal} were satisfactory in the determination of the self-shielding factors when the average chord lengths are calculated from our derived formulae in Table \ref{Tab:Comparisonofl}. The analytical formulae enable its implantation the longer-term application in the analysis of neutron activation and neutron-induced effects of materials for different materials in different geometries, especially neutron shields, using integral parameter representation, instead of spectroscopic one}.

\appendix

\section{Advantage compared to Mont\'{e} Carlo methods}
\label{App:Merit}

The use of Mont\'{e} Carlo simulation software (MC) for calculating the self-shielding factors is feasible but not yet efficacious. The principle underlying MC is to avoid the direct analytical solution of the problem. The goal of MC is to simulate and average a sufficiently large number of particle histories to obtain estimates of the flux which include rigorous approximations. According to Larson, MC of difficult problems are often very costly to set up and run. To make the MC code run with acceptable efficiency, the code users must specify a large number of biasing parameters, which are specialized to each different problem. Determining these parameters can be difficult and time-consuming. Also, even when the biasing parameters are well-chosen, MC converges slowly and non-monotonically with increasing run time. Thus, while MC solutions are free of truncation errors, they are certainly not free of statistical errors, and it is challenging to obtain MC solutions with sufficiently small statistical errors, and with acceptable cost. Finally, the non-analog techniques that have been developed for making MC simulations acceptably efficient and were useful for source-detector problems – in which a detector response in a small portion of phase space is desired –are not useful for obtaining efficient global solutions, over all of phase space. Generally, MC solutions work best when very limited information about the flux (e.g. a single detector response) is desired in a given simulation. MC is feasible for calculating self-shielding of a single sample, it requires time and an experienced user to make the acceptable accuracy and efficiency, the cost most scientists cannot afford to just calculate a single parameter in their routine work. For example, considering set of different samples need to be analyzed using neutron activation for the purpose of elemental analysis, MC SIM needs experience and a lot of time to reduce fluctuation, adopt the geometry, consideration of the neutron transport inside and outside the sample, and the absorption and scattering phenomenon as a function of neutron energy.

Although our intention while deriving and validating the present mathematical formulae was focused on avoiding such costs and enhancing present existing empirical equations and transforming all the problems from the spectroscopic set of parameters, such as thermal cross-section, thickness, width, height, radius, shape, a width of neutron first resonance, the width of the first gamma resonance, etc. into an integrated set of well-known parameters, thermal cross-section,  resonance integral, average chord length.  All remaining factors are calculated from these three parameters. Of course, the thermal cross-section and sample mass and composition are common.

\section{Contribution of velocity distribution}
\label{ASec:Contributionofvelocitydistribution}
According to the results of  Blaauw \cite{Blaauw1995403B}, calculation of the reaction rate is need to be with neutron density averaged macroscopic cross-section (function of velocity) instead of flux averaged macroscopic cross-section (energy dependent). Blaauw found that the self-shielding factors calculated for mono energetic neutrons yields the same results as if they are used with the flux averaged macroscopic cross-section provided that the neutron density averaged macroscopic cross-section  given by
\begin{equation}\label{AEq.BlaauwVelocity} <\Sigma>=\frac{2}{\sqrt{\pi}}\sqrt{\frac{T_o}{T}}\Sigma_o, \end{equation}
is used  instead of the flux averaged capture cross-section given by
\begin{equation}\label{AEq.BlaauwEbnergy} <\Sigma>=\frac{\sqrt{\pi}}{2}\sqrt{\frac{T_o}{T}}\Sigma_o \end{equation}
\noindent  Blaauw \cite{Blaauw1995403B} results showed that the volume-averaged attenuation self-shielding factor in extended neutron distributions, has an extra term that depends on the statistical moments of deviation in reciprocal velocity average. The contribution of this extra factor had been estimated by  Goncalves et al. \cite{Goncalves2004186} to be around 6$\pm$1\%.

The higher~order~terms in Eq. \ref{AEq:SeflShieldEnergyDomain0} (Denoted $\mathcal{R}$ ) can be obtained from  Blaauw \cite{Blaauw1995403B} and adopted to our notions as;
\begin{equation}\label{AEq:RvelocityDistribution}
\mathcal{R}\cong
\sum_{i=1}^{\infty} \frac{(-1)^i}{i!}(\Sigma_a)^i v_o^i\left(\left\langle\frac{1}{v^i}\right\rangle-\left\langle\frac{1}{v}\right\rangle^i\right) \frac{\int_{V}(\vec{r})^i d\vec{r}}{V}
\end{equation}
For the first term,  $i=1$, the value of  $\left(\left\langle\frac{1}{v}\right\rangle-\left\langle\frac{1}{v}\right\rangle\right) $ vanishes. While for approximate spherical symmetry the integral yields the average squared length over volume of sphere, The first term comprises;
\begin{eqnarray}\label{AEq:integration1}
  \int_{V}(\vec{r})^2 d\vec{r}&=&\int_0^{\bar{\ell}}\int_0^{2\pi}\int_0^\pi \underline{r}^2  (\underline{r}\sin\underline{\phi})d\underline{\theta} d\underline{\phi} \underline{r} d\underline{\theta} dr \\&=&\frac{4}{5}\pi \bar{\ell}^5 = \frac{4}{5}\frac{3}{4}\frac{4}{3}\pi \bar{\ell}^3 \bar{\ell}^2 =\frac{3}{5}V \bar{\ell}^2
\end{eqnarray}

For the first approximation, only term of i=2 has an effective contribution. Hence,
\begin{equation}\label{AEq:RvelocityDistribution2}
\mathcal{R}\cong  \frac{+1}{2}(\Sigma_a)^2 v_o^2\left(\left\langle\frac{1}{v^2}\right\rangle-\left\langle\frac{1}{v}\right\rangle^2\right) \frac{3}{5}\bar{\ell}^2
\end{equation}
\noindent For Maxwellian velocity distribution
\begin{eqnarray}\label{AEq:Velocities}
     \left\langle \frac{1}{v}\right\rangle & =&   \frac{2}{\sqrt {\pi }}\frac{1}{v_{o}}\\
      \left\langle \frac{1}{v^2}\right\rangle & =&   2 \frac{1}{v_{o}^2}
\end{eqnarray}

\begin{eqnarray}\label{AEq:RvelocityDistribution3}
  \nonumber \mathcal{R}& \cong (\Sigma_a \bar{\ell})^2 \frac{3}{5}\left(1- \frac{2}{\pi}\right)\approx 0.218 (\Sigma_a \bar{\ell})^2, \,\,\,  Maxwellian & \\
  \nonumber&\cong 0,  \,\,\,  monoenergetic &\\
  &\cong 0,  \,\,\,  20v_o\leq v\leq4\times10^7v_o &
\end{eqnarray}
The first formulae is valid only for the entire range of Maxwellian neutron distribution. However, for practical use, only the epi-thermal range between cadmium cutoff at 0.5 eV upto about 1 MeV is used. In such cases,  the difference $\left(\left\langle\frac{1}{v^2}\right\rangle-\left\langle\frac{1}{v}\right\rangle^2\right)$ practically vanishes,  $\mathcal{R}\cong 0 $.

There is an additional reason why this value is ignored within the sample in the present work. The Blaauw \cite{Blaauw1995403B} derivation is based on the idea that the neutron flux distribution have constant shape as it passes through the depth $\bar{\ell}$. However, there is a simple experimental remark  in neutron physics: whenever the neutron energy distribution repudiate the proper thermalization distribution (Maxwellian+1/E dependance in epi-thermal region) by any mean such as absorption, the neutrons rapidly redistribute its velocity population within the diffusion distance to follow the proper distribution -- up to  thermalization. The idea is, even if there is absorption of neutrons having a velocity $v$ in Eq. \ref{AEq:RvelocityDistribution} at some distance, there were a sort of recovery of that distribution. And hence, the difference in Eq. \ref{AEq:RvelocityDistribution2} has much less value than expected by Blaauw.

\section{Determination of chord lengths based on neutron transport formulae}
\label{AApp:Derivation}

The time dependent diffusion equation comprises;
\begin{equation}\label{AEq:TimeDepDiffusion}
  \frac{1}{{\mathrm{v}_{av}}}\frac{\partial \varphi (\vec{r},t)}{\partial t}=\nabla \cdot J(\vec{r},t)-{{\Sigma }_{a}}\varphi (\vec{r},t)+Q(\vec{r},t),
\end{equation}
\noindent where   $J(\vec{r},t)=$ $D(\vec{r},t)$$\nabla \varphi (\vec{r},t)$  is the  neutron current,  $Q(\vec{r},t)$ is the neutron production rate within the medium in units of n~cm$^{-3}$s$^{-1}$. Under the condition of steady-state ($\frac{\partial \varphi (\vec{r},t)}{\partial t}=0$) and Considering:
 \begin{itemize}
    \item the sample is embedded within a uniform neutron field in which the flux outside it, $\varphi_o$, to be isotropic,  uniform and does not depend on the diffusion within the sample,
  \item our situation of sample absorbing neutrons not generating it,
 \end{itemize}
the solution of the problem comes as\emph{ difference-problem} in the steady-state where  $Q(\vec{r},t)$ equated to $\varphi_o$,  and considering only the difference replacing $\varphi(\vec{r})$ by $\varphi_o-phi(\vec{r})$: Then

\begin{equation}\label{AEq:Diffusion}
  \nabla \cdot D(\vec{r})\nabla \phi(\vec{r}) -\Sigma_a(\vec{r})\phi(\vec{r})=0
\end{equation}

In the homogenous isotropic medium, $D(\vec{r})$ and $\Sigma_a(\vec{r})$ are constants.
\begin{equation}\label{AEq:leakagerate1}
    \nabla^2\phi(\vec{r})-B^2\phi(\vec{r})=0,
\end{equation}
The $B$ factor is  the geometric buckling factor in reactor physics. Taking into consideration that $\Sigma_a=$$1/\lambda_a$, and $D=\lambda_{tr}/3$ where $\lambda_{a}$ is the absorption mean-free path and $\lambda_{tr}$ is the transport scattering diffusion length given by the  more advanced transport theory in terms of transport and absorption cross-sections equation as \cite{Espinosaparedes20081963,Tzika2004177};
\begin{equation}\label{AEq:DiffusionCoefficient}
  \lambda_{tr}=\frac{1}{\Sigma_a+\Sigma_s(1-\bar{\mu})},
\end{equation}
\noindent where $\bar{\mu}=\frac{2}{3A}$ is average value of the cosine of the angle in the lab system. So,
\begin{equation}\label{AEq:B2}
  B^2=\frac{\Sigma_a [cm^{-1}]}{D [cm\, s^{-1}]},
\end{equation}

\subsection{Rectangular geometries}
\label{AApp:DerivationCartCoor}
In cartesian coordinates, Eq. \ref{AEq:leakagerate1} is reduced to three independent equations by separation of variables assuming $\phi(\underline{x},\underline{y},\underline{z})=$ $\phi_x(\underline{x})$ $\phi_y(\underline{y})$ $\phi_z(\underline{z})$. I.e.
\begin{eqnarray}\label{AAEq:leakagerate0}
\left(\frac{d^2 }{d \underline{x}^2} -B_1^2\right)\phi_x(\underline{x})&=&0\\
\left(\frac{d^2 }{d \underline{y}^2}-B_2^2\right)\phi_y(\underline{y})&=&0\\
\left(\frac{d^2 }{d \underline{z}^2}-B_3^2\right)\phi_z(\underline{z})&=&0,
\end{eqnarray}
The general solution is:
\begin{equation}\label{AAEq:Leakagerate2}
   \phi(\underline{x},\underline{y},\underline{z})=\mathrm{Const.} e^{-iB_1\,\underline{x}}e^{-iB_2\,\underline{y}}e^{-iB_3\,\underline{z}}
\end{equation}

\subsection{Cylindrical geometries}
\label{AApp:DerivationCylinCoor}

For  a definite convex shapes of having cylindrical geometries, Eq. \ref{AEq:leakagerate1} becomes the  Helmholtz differential equation;
\begin{eqnarray}\label{AAEq:leakagerate2}
   & \left(\frac{1}{\underline{\rho}}\frac{\partial}{\partial \underline{\rho}}\left(\underline{\rho} \frac{\partial}{\partial \underline{\rho}}\right) \text{+} \frac{1}{\underline{\rho}^2}\frac{\partial^2}{\partial \underline{\phi}^2} \text{+} \frac{\partial^2 }{\partial \underline{z}^2}   \text{+}  B^2 \right) \phi(\underline{\rho},\underline{\phi},\underline{z}) =0.&
\end{eqnarray}
In which the metric tensor scale factors are 1, $\rho$, and 1 for the coordinates $\rho$, $\phi$, $z$, respectively. Separation of variables is done by writing
$\phi(\underline{\rho},\underline{\phi},\underline{z}) = \phi_\rho(\underline{\rho})\phi_\phi(\underline{\phi})\phi_z(\underline{z})$. Eq. \ref{AAEq:leakagerate2} becomes;
\begin{eqnarray}\label{AAEq:leakagerate3}
\nonumber & \left({\frac{\underline{\rho}^2}{ \phi_\rho}\frac{d^2\phi_\rho}{d\underline{\rho}^2}+ \frac{\underline{\rho}}{\phi_\rho}\frac{d\phi_\rho}{d\underline{\rho}}}\right) + \frac{1}{\phi_\phi}\frac{d^2 \phi_\phi}{d \underline{\phi}^2} + \frac{\underline{\rho}^2}{\phi_z}\frac{d^2 \phi_z}{d \underline{z}^2}- &\\ & \underline{\rho}^2B^2 =0. &
\end{eqnarray}
Solution requires  negative separation constants -- say $m^2$ in order to maintain the  periodicity in $\varphi$;  hence,
\begin{equation}\label{AAEq:leakagerate4}
\frac{1}{\phi_\phi}\frac{d^2\phi_\phi}{d\underline{\phi}^2}= -m^2,
\end{equation}
which has a general solution
\begin{equation}\label{AAEq:leakagerate5}
\phi(\underline{\phi}) = \mathrm{C}_1\cos{m\,\underline{\phi}}+\mathrm{C}_2\sin{m\,\underline{\phi}}.
\end{equation}
where $C_1$, and $C_2$ are constants. Hence,
\begin{equation}\label{AAEq:leakagerate61}
\left({\frac{1}{\phi_\rho}\frac{d^2\phi_\rho}{d\underline{\rho}^2}+ \frac{1}{\underline{\rho} \phi_\rho}\frac{d\phi_\rho}{d\underline{\rho}}}\right) - \frac{m^2 }{\underline{\rho}^2}+ B^2+ \frac{1}{\phi_z}\frac{d^2 \phi_z}{d \underline{z}^2}=0,
\end{equation}
i.e. for finite cylinder, there are two coordinate lengths, radius and hight
\begin{eqnarray}\label{AAEq:leakagerate6}
\nonumber& \left({\frac{1}{\phi_\rho}\frac{d^2\phi_\rho}{d\underline{\rho}^2}+ \frac{1}{\underline{\rho} \phi_\rho}\frac{d\phi_\rho}{d\underline{\rho}}}\right) - \frac{m^2 }{\underline{\rho}^2}+& \\ & B_1^2+B_2^2+ \frac{1}{\phi_z}\frac{d^2 \phi_z }{d\underline{z}^2}=0&
\end{eqnarray}
The $\phi_z$ must not be sinusoidal at $\pm \infty$  which lead to positive separation constant -- say $n^2$;
\begin{equation}\label{AAEq:leakagerate7}
\frac{1}{\phi_z}\frac{d^2\phi_z}{d\underline{z}^2}=  n^2-B_2^2,
\end{equation}
\begin{equation}\label{AAEq:leakagerate8}
\frac{d^2\phi_\rho}{d\underline{\rho}^2}+ \frac{1}{\underline{\rho}}\frac{d\phi_\rho}{d\underline{\rho}}+ \left({n^2 +B_1^2- \frac{m^2}{\underline{\rho}^2}}\right)\phi_\rho = 0.
\end{equation}
The solutions are
\begin{eqnarray}\label{AAEq:leakagerate9}
\nonumber & \phi_Z(\underline{z}) = \mathrm{C}_3e^{-\sqrt{n^2-B_2^2}\underline{z}}+\mathrm{C}_4e^{\sqrt{n^2-B_2^2}\underline{z}} & \\ & \, if \, n^2>B_2^2 & \\
\nonumber & = \mathrm{C}_5\cos{\sqrt{B_2^2-n^2}\underline{z}}+\mathrm{C}_6\sin{\sqrt{B_2^2-n^2}\underline{z}}  & \\  &\, if \, n^2\leq B_2^2 &
\end{eqnarray}
\begin{eqnarray}\label{AAEq:leakagerate10}
&\phi_\rho(\underline{\rho}) =\mathrm{C}_7J_m(\sqrt{n^2\text{+}B_1^2}\underline{\rho})\text{+}\mathrm{C}_8Y_m(\sqrt{n^2\text{+}B_1^2}\underline{\rho}),&
\end{eqnarray}
where $J_m$ and $Y_m$ are Bessel functions of the first kind and second Kind, respectively. These results requires that $n$ and $m$ be integers.  $Y_n(0)$=$-\infty$ which lead to un-physical solution, hence, $\mathrm{C}_2$=0 and $\mathrm{C}_8$=0. Similarly, $\mathrm{C}_4$=0 and $\mathrm{C}_6$. The solution is reduced to;
\begin{eqnarray}\label{AAEq:leakagerate11}
   \nonumber  & {\phi}(\underline{\rho},\underline{\phi},\underline{z}) =\sum_{m=0}^{\infty}\sum_{n=0}^{\infty}C_{mn} J_m (\sqrt{n^2+B_1^2}\underline{\rho})  \cos{m\,\underline{\phi}} & \\                                                         &e^{-\sqrt{n^2-B_2^2}\underline{z}}  \, if \,n^2>B_2^2 &\\
 \nonumber &=\sum_{m=0}^{\infty}\sum_{n=0}^{\infty}C_{mn} J_m (\sqrt{n^2+B_1^2}\underline{\rho})  \cos{m\,\underline{\phi}} & \\ & \cos{\sqrt{B_2^2-n^2}\underline{z}} \, if \, n^2\leq B_2^2 &
\end{eqnarray}
where $C_{mn}$ is a constant that depends on the values of m and n.

\subsection{Spherical geometries}
\label{AApp:DerivationSpherCoor}

In spherical coordinates, Eq. \ref{AEq:leakagerate1} resembles;
\begin{eqnarray}\label{AAEq:leakagerate13}
\nonumber & \left( \frac{1}{\underline{r}^{2}}\frac{\partial}{\partial \underline{r}}\left(\underline{r}^{2}\frac{\partial}{\partial \underline{r}}\right)+\frac{1}{ \underline{r}^{2}\sin \underline{\theta} }\frac{\partial}{\partial \underline{\theta} }\left(\sin \underline{\theta} \frac{\partial}{\partial \underline{\theta} }\right)+\right. & \\
& \left.\frac{1}{\underline{r}^{2}\!\sin ^{2}\underline{\theta} }\frac{\partial ^{2}}{\partial \underline{\phi} ^{2}}\right) \phi +B_1^2\phi=0. &
\end{eqnarray}
Because of the spherical symmetry there is only one value of $B$=$B_1$. Values of $B_2$ and $B_3$ vanishes; i.e. $\ell_2=\infty$ and $\ell_3=\infty$.
\begin{eqnarray}\label{AAEq:leakagerate14}
\nonumber & \left( \frac{1}{\phi_{r}}\frac{d }{ d r}\left(\underline{r}^{2}\frac{d \phi_r }{ d \underline{r}}\right)+\frac{1 }{ \phi_\theta}\frac{1 }{ \sin \underline{\theta} }\frac{d }{ d \underline{\theta} }\left(\sin \underline{\theta} \frac{d \phi_\theta}{ d \underline{\theta} }\right)+\right. & \\ & \left. \frac{1}{ \phi_\phi}\frac{1 }{ \sin ^{2}\underline{\theta} }\frac{d ^{2} \phi_\phi}{d \underline{\phi} ^{2}}\right)  +\underline{r}^2B_1^2=0.&
\end{eqnarray}
 Separation of variable requires substitution of $\phi(\underline{r}, \underline{\theta},\underline{\phi}) $ by  $\phi_r(\underline{r})\phi_\theta(\underline{\theta})\phi_\phi(\underline{\phi})$. Separating the $r$ term with separation constant $l(l+1)$
\begin{equation}\label{AAEq:leakagerate15}
 \frac{1 }{\phi_r} \frac{d }{d \underline{r}}\left(\underline{r}^{2}\frac{d \phi_r }{ d \underline{r}}\right)-l(l+1)+\underline{r}^2B_1^2=0
\end{equation}
\begin{eqnarray}\label{AAEq:leakagerate16}
\nonumber & \frac{1 }{ \phi_\theta}\frac{1 }{ \sin \underline{\theta} }\frac{d }{ d \underline{\theta} }\left(\sin \underline{\theta} \frac{d \phi_\theta}{ d \underline{\theta} }\right)+\frac{1 }{ \phi_\varphi}\frac{1 }{ \sin ^{2}\underline{\theta} }\frac{d ^{2} \phi_\varphi}{ d \underline{\varphi} ^{2}} & \\ & + l(l+1)=0&\end{eqnarray}
Eq. \ref{AAEq:leakagerate16} is separated by separation constant $m^2$, then
\begin{eqnarray}\label{AAEq:leakagerate17}
 & \frac{1 }{ \phi_\theta}\frac{1 }{ \sin \underline{\theta} }\frac{d }{ d \underline{\theta} }\left(\sin \underline{\theta} \frac{d \phi_\theta}{ d \underline{\theta} }\right)  +l(l+1)\sin\underline{\theta} -m^2=0 &\end{eqnarray}
\begin{equation}\label{AAEq:leakagerate18}
\frac{1 }{ \phi_\phi}\frac{d ^{2} \phi_\phi}{ d \underline{\phi} ^{2}} +m^2=0\end{equation}
solution of Eqs. \ref{AAEq:leakagerate15}, \ref{AAEq:leakagerate17}, and \ref{AAEq:leakagerate18} yield the following solution;
\begin{equation}\label{AAEq:leakagerate19}
  \phi(\underline{r},\underline{\theta},\underline{\phi})= \sum_{l=0}^{\infty}\sum_{m=0}^{\infty}C_{mn} \;j_l(B_1 \,\underline{r}) \; P_l^m(\cos\underline{\theta}) \cos{m\,\underline{\phi}}
\end{equation}
\noindent by ignoring the anti symmetric terms. Here, $P_l^m(\cos\underline{\theta})$ gives the associated Legendre polynomial while $j_l(B \,\underline{r})$ is the spherical Bessel function.

\section{Comparison with empirical formula}
\label{Asec:Comparison}
Figures \ref{AFig:Comparisonwire} and \ref{AFig:Comparisonfoils} represent the thermal self-shielding factor calculated using empirical equations given by Refs. \cite{Gonalves2001447,Martinho2003371,Martinho2004637,Goncalves2004186,Salgado2004426}.
\begin{figure}[htb]
  \centering
  \includegraphics[width=0.95\linewidth]{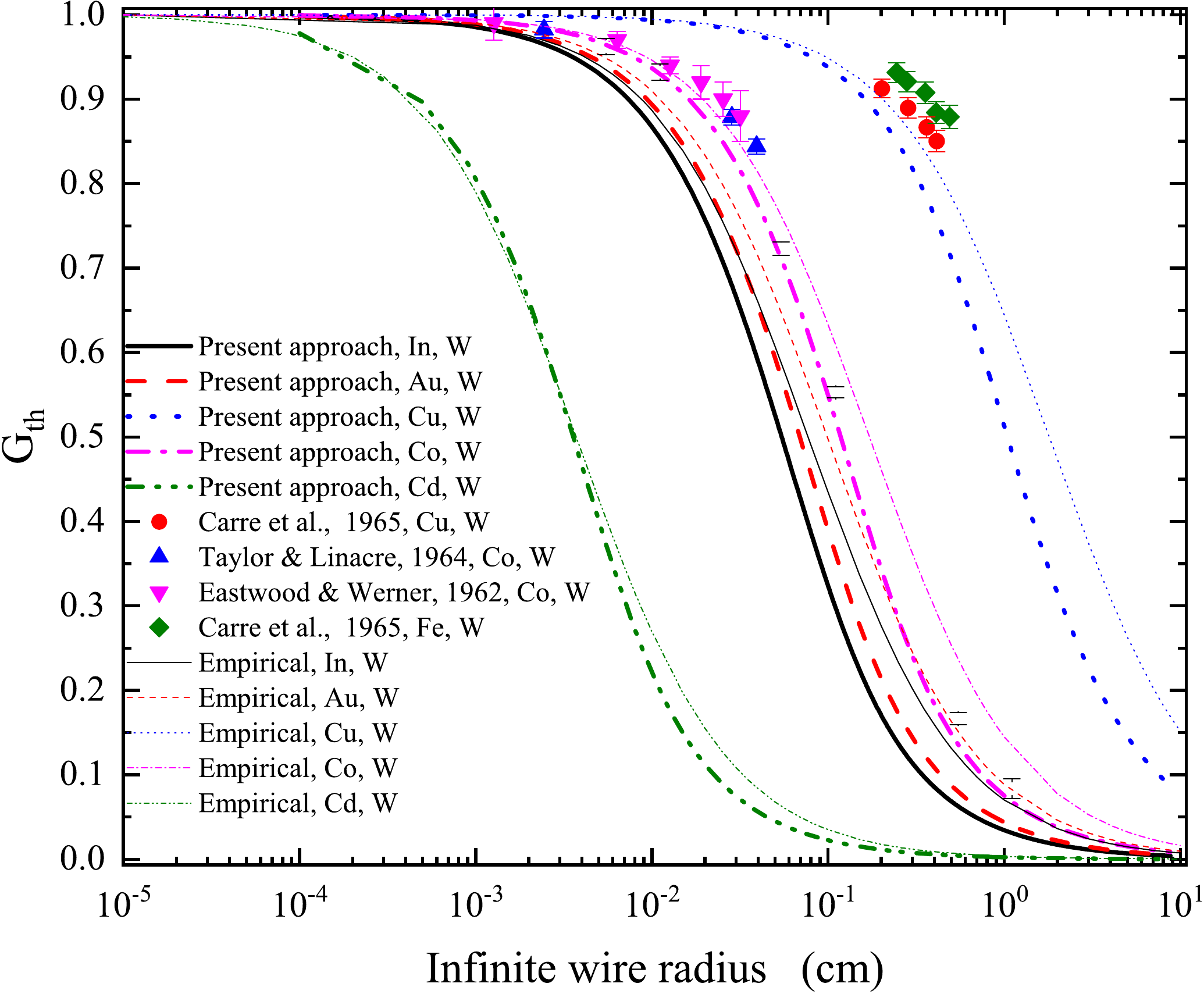}
  \caption{A comparison between the present general approach of \emph{ab initio} calculations, considering the extreme cases of infinite  wire, and the empirical equation given in Refs. \cite{Gonalves2001447,Martinho2003371,Martinho2004637,Goncalves2004186,Salgado2004426}. Experimental data from Taylor $\&$ Linacre \cite{taylor1964use}, Carre et al. \cite{carre1965etudes}, were digitized from Martinho et al. \cite{Martinho2004637}. While, the data of Eastwood $\&$ Werner \cite{eastwood1962resonance} for Co Wire was collected from their original values.}\label{AFig:Comparisonwire}
\end{figure}
\begin{figure}[htb]
  \centering
  \includegraphics[width=0.95\linewidth]{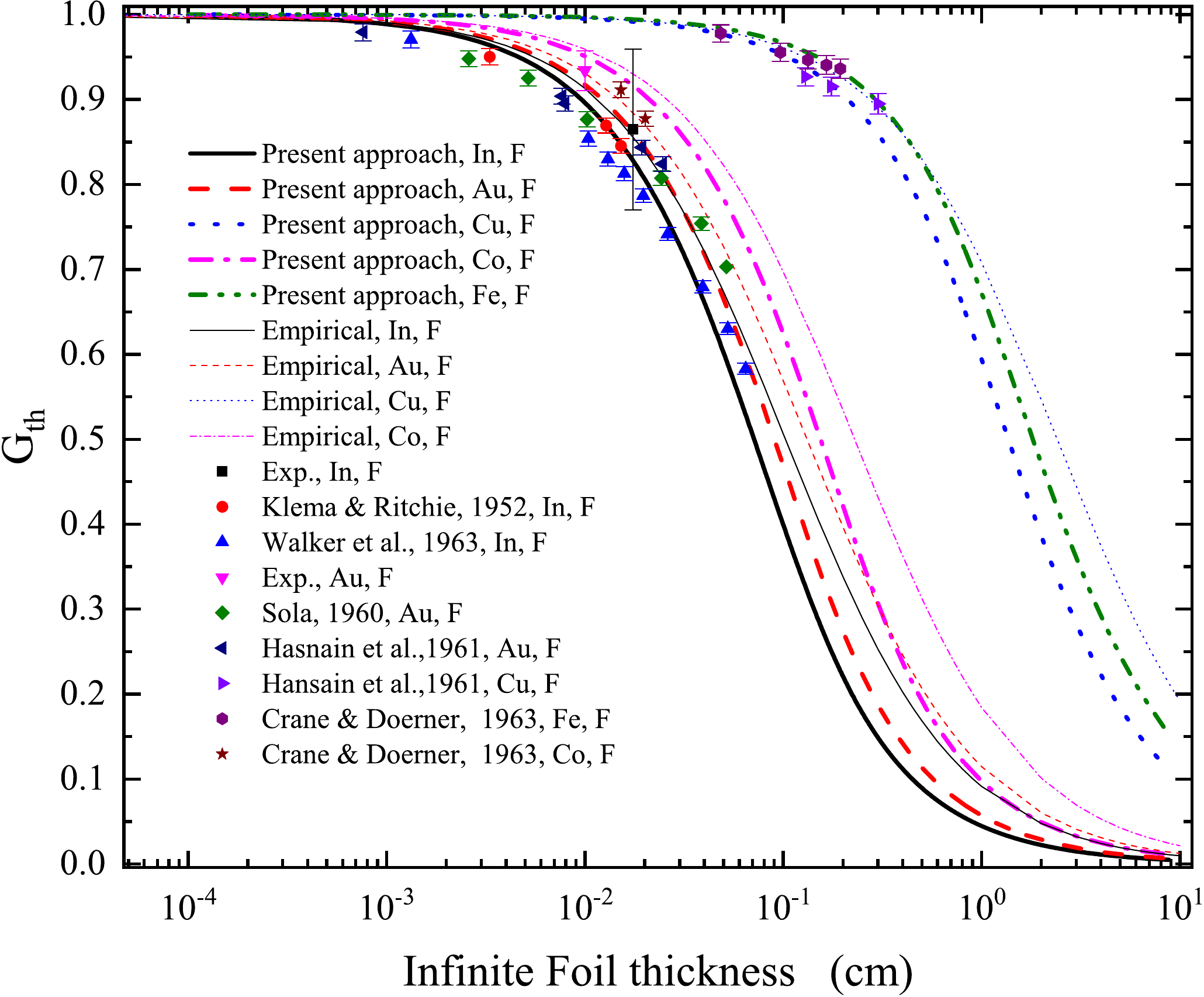}
  \caption{A comparison between the present approach of \emph{ab initio} calculations, considering the extreme cases of infinite  foil, and the empirical equation given in Refs. \cite{Gonalves2001447,Martinho2003371,Martinho2004637,Goncalves2004186,Salgado2004426}. Experimental data from Hasnain et al. \cite{hasnain1961thermal}, Sola \cite{Sola1960},  Walker et al. \cite{walker1963thermal}, Klema \cite{klema1952thermal}, and Crane $\&$ Doerner \cite{crane1963thermal} were digitized from Martinho et al. \cite{Martinho2004637}.}\label{AFig:Comparisonfoils}
\end{figure}

 Their curves were calculated with the specific values of cross-section given in their manuscripts, which does not equal to the recommended cross-sections in literatures. Our approach was calculated using the extreme approximation of infinite foil (having only one variable with is the thickness) and with the general cross-section values in Refs. \cite{Sukhoruchkin1998book,Sukhoruchkin2009NeutResPara,Elmaghraby2016Shape}.
Based on this comparison based on extreme cases of infinite wire and infinite foil, the results of our approach matched the empirical equation in most cases, where it was already succeeded. Note that: there is no such infinite foil or infinite wire in experimental situations.

\section*{Conflict of interest}
  The authors declare that they have no known source for conflict of interest with any person.

\newpage


%

\newpage

~
\newpage

\begin{widetext}
\begin{center}

\textbf{Graphical Abstract}

  \includegraphics[width=\textwidth]{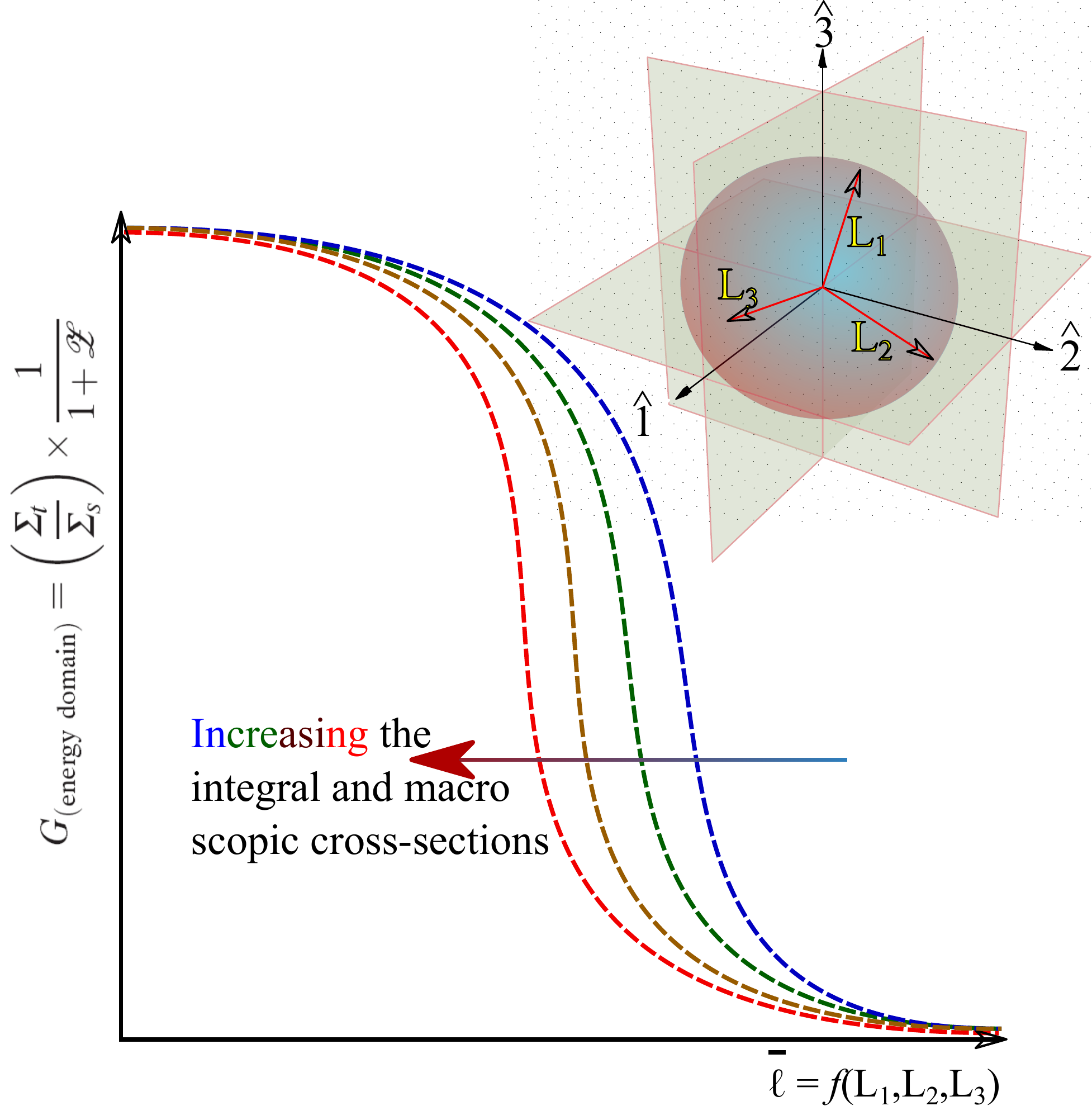}

\end{center}
\end{widetext}

\end{document}